# Structure and dielectric properties of (x)BaFe$_{0.5}$Nb$_{0.5}$O$_3$ – (1-x)KNbO$_3$ solid solutions synthesized through solution combustion route

Vijay Khopkar and Balaram Sahoo[†]

*Materials Research Centre, Indian Institute of Science, Bangalore 560012 India*

[†]Corresponding author: bsahoo@iisc.ac.in (B. Sahoo)

Ph: +91 8022932943

# Abstract

We demonstrate that the solution combustion reaction (SCR) route is suitable for the synthesis of phase pure $(x)BaFe_{0.5}Nb_{0.5}O_3$-$(1-x)KNbO_3$ (x = 0, 0.2, 0.4, 0.6, 0.8, 1) (BFN-KN) solid solutions due to atomic level of mixing of precursors than that of the solid-state reaction (SSR) route. A variation in composition 'x' of our double perovskite samples leads to a gradual variation in the structural disorder associated with the disorder in the distribution in the cationic positions, the lattice strain, and the defects at the grain boundaries. With the increase in the BFN content in the solid solutions, a decrease in bandgap and a corresponding change in the color of the samples are observed. Furthermore, three distinct characteristic features in the frequency-dependent polarization behavior in our samples are observed at different frequencies. These features observed to be sustaining up to gradually increasing frequencies are assigned to (1) the interfacial polarization originating due to the structural disorder present at the grain boundaries, (2) the cationic positions leading to tilting/expansion of oxygen octahedra, and (3) the ionic size mismatch leading to strain distribution. The results are well supported by the results obtained for the microstructural, structural, and optical properties of our samples. Hence, our work provides a better understanding of the dielectric polarization behavior of double perovskites.

## 1. Introduction

Materials with a high dielectric constant are useful for applications in many electronic devices such as mechanical sensors, memory devices, actuators, gas sensors, energy storage devices, etc. [1, 2]. Due to better dielectric and functional properties, lead-based $ABO_3$ type perovskite structured materials were most commonly looked at [3–6], but the toxic nature of lead (Pb) diverts the attention of the scientist to search for alternate lead-free perovskite materials [7, 8]. In this regard, many perovskites, such as $BaTiO_3$, $BiFeO_3$, $KNbO_3$, etc., are the material of interest [9]. Among them, potassium niobate (KN), $KNbO_3$ is the well-known classical ferroelectric material having orthorhombic crystal structure at RT [10]. This material shows a high dielectric constant value only at the Curie temperature (434 ºC) [11]. However, for application purposes, it is desirable to maintain the high dielectric constant, not only at the Curie temperature but over a wide range of operating temperatures. Achieving this is the real concern for the present scientific studies. We believe that to achieve this objective, the strategy that can be used effectively is by introducing a disorder in the occupancy of the cations in the crystal structure of the materials. We expect that with an increase in cationic position and size and charge disorder at A and B sites, the material would begin to exhibit a diffuse phase transition, and the dielectric constant would remain high over a wide range of temperatures [12, 13]. Previously, this aspect is studied in various binary solid solutions for improving the dielectric properties, e.g., in classic ferroelectric materials [14, 15].

As far as dielectric properties are concerned, the leakage current is the detrimental factor. The leakage current arises due to the presence of phase impurities, defects, and disorders in the samples. Hence, the phase pure ceramic samples are very important to demonstrate their functional properties for various applications. As impurities hinder the actual properties of parent compounds, the samples need to be pure at an atomic level. Traditionally, most of the ceramic compounds and their solid solutions are synthesized by conventional solid-state reaction (SSR) route[16–18], although other methods, like sol-gel [19], co-precipitation [20, 21], mechanical alloying [22, 23], hydrothermal [24–26], solution combustion [27, 28] etc. are also used. In this SSR method, carbonated and oxides precursors are generally used in the solid/powder form with micron-sized particles. During the initial stage, an atomic level mixing of these solid precursors is not possible, and in the subsequent stages of calcination, grinding, and sintering, an atomic level of homogeneity is also not achievable. Furthermore, to facilitate better diffusion during the sintering process, the samples need to be sintered for a long time, which is time and energy-consuming, and often not feasible. Hence, this repeated grinding and sintering continues until one does not observe any detectable impurity phase by techniques like XRD. However, a sub-nanometric level detection of the impurity phases, along with the required ceramic phase, is often not possible by XRD. So, to synthesize the ceramic sample by alternate methods needs to be used, where atomic level mixing can be achieved at the precursor level, such as in solution combustion reaction (SCR) synthesis [29–34]. The precursors used in this method are mostly metal-nitrates, along with a suitable fuel, which is water-soluble, leading to atomic/molecular level mixing. A further advantage of this SCR method is the reduction in the required reaction temperature (down to the flash temperature of the fuel only!)) [35–37]. This saves the energy and the reaction time to synthesize the ceramic samples. Due to all these advantages, the solution combustion route is far more efficient than the conventional solid-state reaction route, and hence, it is our chosen method for the synthesis of the samples.

Among the lead-free functional ceramic material, the double perovskite like barium iron niobate (BFN, $BaFe_{0.5}Nb_{0.5}O_3$) has attracted wide potential interest because it shows a large dielectric constant value over wide temperature range [17, 38, 39]. Furthermore, as BFN is a lead-free double perovskite material having a mixture of two different cations ($Fe^{3+}$ and $Nb^{5+}$) at the B site, this material is explored as a relaxor ferroelectric, which may show a higher dielectric constant over a wide range of temperatures; however, this property of BFN is under debate [40–45]. As BFN has a pseudo-cubic structure[46, 47], the origin of ferroelectric behavior at RT needs a better investigation. As stated earlier, KN is a ferroelectric material with a perovskite structure. Hence, to extend the high dielectric constant value of KN over a wide temperature range, it is interesting to make a solid solution with BFN. The synthesis and dielectric study of the solid solution of BFN and KN is adopted in this work. We expect to understand the dielectric polarization behavior of BFN and its origin through this study. As per our knowledge, only a few reports are available with only minor exploration of the dielectric behavior of BFN-KN solid solution or any similar solid solutions [48, 49]. These reports only include a brief structural and optical property study of the solid solution system for solar cell application; however, a detailed study on the dielectric behavior of BFN-KN solid solution lacks in the literature. In this work, we investigated the dielectric property of this solid solution along with its structural and optical properties.

## 2. Experimental details

We have used two different methods, solid-state reaction (SSR) method and the solution combustion reaction (SCR) method, for the synthesis of (x) $BaFe_{0.5}Nb_{0.5}O_3$ - (1-x)

KNbO$_3$(x = 0, 0.2, 0.4, 0.6, 0.8, 1) solid solutions. For the solid-state reaction (SSR) method, precursors such as BaCO$_3$ (99%, Merck), α-Fe$_2$O$_3$ (95%, SDFCL), Na$_2$O$_5$ (98%, Merck), and K$_2$CO$_3$ (99% Sigma-Aldrich) were used in the stoichiometric amount, ground thoroughly for mixing, and the mixture was calcined between 800-1200 °C for 4-10 hours using alumina crucible in the muffle furnace. The temperature of the mixture was gradually increased during the attempt to obtain phase pure samples. The structure of the resulting solid solutions, along with their phase purity, was studied through the analysis of the powder X-ray diffraction and the Mössbauer spectroscopy results.

For the synthesis of the BFN-KN solid solutions by solution combustion reaction (SCR) method, the nitrate precursors such as Ba(NO$_3$)$_2$, KNO$_3$, Fe(NO$_3$)$_3$.9H$_2$O, and Na$_2$O$_5$ were used (known as oxidizers) along with glycine as a fuel. The burning of the fuel provides the required heat for the formation of the compounds. The stoichiometric amounts of the fuel and oxidizers were determined according to the propellant chemistry [29, 50, 51]. The stoichiometric amounts of precursors were dissolved in a minimum amount of deionized water with constant stirring to form a homogenous solution and kept in a preheated furnace at 550 °C. The solution starts boiling, and after few minutes, the combustion reaction starts. After the combustion reaction, the obtained coarse powders were ground thoroughly and characterized. The samples were further calcined sequentially between 800-1200 °C in an alumina crucible for 2-4 hours until pure phase solid solution was obtained. The samples were then pressed into the disc-shaped pellets of 1 cm diameter with an optimized uniaxial load (of 25 kN) and sintered between 1000-1250 °C for 2 hours to get the phase pure dense pellets. All the characterizations were performed for these (phase pure) samples.

The powder X-ray diffraction (XRD) patterns of all the crushed pellet samples were obtained by using a 'PANalytical X'PertPro' diffractometer (Cu K$_\alpha$ radiation (λ = 1.54 Å) and Ni filter) with the 2θ range of 10 - 90° and a step size of 0.0263°. The Mössbauer spectra were recorded by $^{57}$Fe-Mössbauer spectroscopy using a SEE Co. (USA) spectrometer. The scanning electron microscopy (SEM) images were obtained through ′ESEM Quanta′ instrument at an operating voltage of 10 kV. The UV-Vis diffuse reflectance spectra were measured using the LAMBA 750 UV/Vis/NIR spectrometer from PerkinElmer with deuterium and tungsten halogen lamp as the light source and a PbS detector. The Raman spectrum of pellets was recorded using a 532 nm excitation laser source on LabRam HR, HORIBA equipment. The room temperature dielectric measurement was performed using precision impedance analyzer 4294A Agilent in a frequency range of 40 Hz to 5 MHz with an applied voltage of 500 mV.

## 3. Results and Discussion

### 3.1 X-ray Diffraction

The evolution in phase purity of the BFN and KN samples, along with that of their solid solutions synthesized via the SSR and SCR route followed by calcination at gradually increasing temperatures between 800 – 1200 °C are investigated through the analysis of their XRD patterns. Note that the crystal structure of BFN and KN is pseudo-cubic (space group: *Pm-3m*) and orthorhombic (space group: *Amm2*), respectively [19, 39]. Fig. 1(a) shows the XRD patterns of the selected SSR samples having the best phase purity. Clearly, the samples with x=0.4, 0.6 and 0.8, still show a minute contribution from the impurity phases, with the impurity peaks marked as '*' in Fig. 1(a). Beyond this calcination temperature, the phase purity of the samples worsens. This could be understood by the growth of the impurity phases as the calcination temperature increases[29]. Furthermore, it can be seen in Fig. 1(a) that for obtaining pure phase KN and BFN samples, calcination temperatures of ~ 800 °C and 1200 °C are required, which matches with other literature reports[7, 39]. According to our observation, by increasing the calcination temperature for the KN sample, impurity phases start forming, indicating the evaporation of potassium from the KN sample [7]. On the other hand, any temperatures lower than 1200 °C are not sufficient to form phase-pure BFN. Hence, it can be rationalized that the synthesis of BFN-KN solid solution of any composition is not so easy due to the volatility of K above ~ 800-850 °C and the simultaneous inability of lower temperatures below 1200 °C to form the phase-pure BFN. Hence, as the content of BFN in the solid solution increases, the required calcination temperature for obtaining the best samples also increases, but for many samples, it is not possible to form phase pure solid solutions. The main reason behind the non-formation of phase pure KN-BFN solid solutions could be the inability of the SSR method to provide an atomic or molecular level of mixing of the precursor. This suggests that for the synthesis of phase pure KN-BFN solid solutions a method that facilitates atomic-level mixing of the precursor is required. As the SCR method could be well suited for this, we have synthesized our BFN-KN solid solution samples through this SCR method, and demonstrated that SCR method is better than SSR method for synthesis of perovskite ceramics, as discussed below.

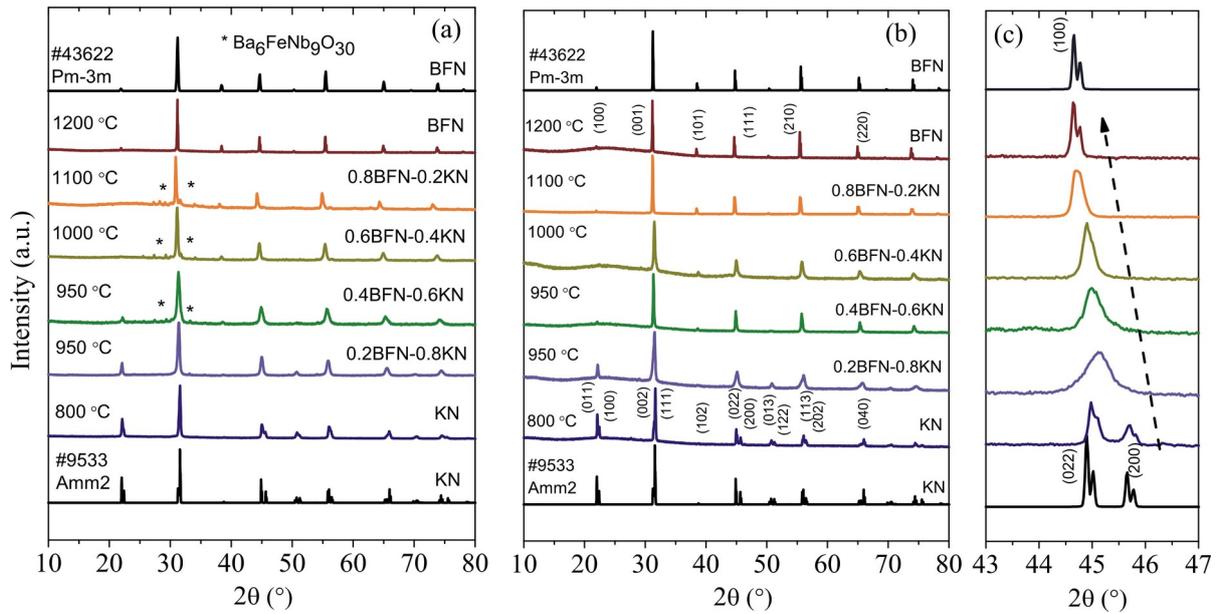

Fig.1. The XRD pattern of the (x) BFN - (1-x) KN (x = 0, 0.2, 0.4, 0.6, 0.8, 1) solid solution samples synthesis by (a) Solid state reaction (SSR) route, and (b) Solution combustion reaction (SCR) route. The optimum calcination temperature for each sample is also mentioned. (c) Zoomed in view of the Bragg peaks positioned at about 2θ = 45°, for all the studied samples synthesized via SCR route.

      The XRD patterns of the SCR synthesized samples recorded after calcination at different temperatures are shown in Fig. S1 (electronic supporting information (ESI) file). It can be observed from Fig. S1 that a gradual evolution in the phase purity of the BFN, KN, and their solid solution samples with an increase in calcination temperature occurs. The XRD patterns of the best samples obtained at the optimum calcination conditions are plotted in Fig. 1(b). There is some similarity in the XRD patterns of the samples prepared by SSR and SCR methods. However, the important differences that stand out for the SCR synthesized samples is that: (1) we obtained phase pure samples by the SCR method for all the compositions of the solid solutions, (2) the calcination temperature required for the formation of the phase pure BFN-KN solid solution is considerably lower than that required for SSR method (but even then, the SSR samples are not phase-pure). Hence, it can be understood that atomic-level mixing of the precursors gives us the handle to obtain phase pure BFN-KN solid solution ceramics at a low temperature by SCR method than that is required for the SSR method. Note here that the reduction in calcination temperature also minimizes the issues related to the volatility of K (Fig. S2 in ESI). Hence, the above XRD results confirm that the SCR method is the suitable method for the synthesis of ceramic compounds and solid

solutions, where there is a considerable mismatch in the phase formation temperatures of the individual compounds.

Furthermore, to understand the evolution of structural details of the solid solutions with the variation of the composition x, in Fig. 1(c), we have shown the zoomed-in view of the XRD peak observed at the 2θ of ~ 45°. It can be identified in Fig. 1(c) that just by the addition of a small amount (x= 0.2) of BFN into KN, the crystal structure changes from orthorhombic (for pure KN) to pseudo-cubic (for BFN). We believe that this occurs due to the larger size of the A-site $Ba^{2+}$ cation, which is dominant in controlling the structure of the overall solid solution. As the BFN content (x) increases further in the solid solutions, the (100) XRD peak shifts toward a lower 2θ value, i.e., the lattice parameter increases, without any change in the pseudo-cubic structure of the samples with compositions up to the pure BFN. The increase in lattice parameters due to the addition of BFN can be understood as due to the introduction of the bigger-sized $Ba^{2+}$ cations at the A-site of the double perovskite structure[52]. The random distribution of the $Ba^{2+}$ and $K^+$ cations at the A-sites and $Fe^{3+}$ and $Nb^{5+}$ cations at the B-sites lead to distribution in local chemical composition and in the inter-atomic distances throughout the matrix of the sample. Please note further that the full width at half maximum (FWHM) of the XRD Bragg-peaks observed in Fig. 1(c) is the lowest for both the pure compounds (KN and BFN), but it is the highest for the solid solution with x =0.2, with a gradual decrease in FWHM with an increase in BFN content x. This variation can be understood as due to the lattice strain developing due to the accommodation of bigger $Ba^{2+}$ cations at the A-site of the BFN-KN solid solutions. Hence, even with the addition of a small amount of BFN into KN, the structure changes with a huge strain broadening. Once the pseudo-cubic structure is established for the solid solution, the strain decreases, and the influence of the smaller $K^+$ cations is not observable. As it will be discussed later, the composition-dependent distribution in local chemical composition, inter-atomic spacing, and the distribution in strain in the solid solutions play a very important role in controlling the dielectric polarization behavior of the solid solution samples.

### 3.2 Mössbauer spectroscopy

Fig. 2 (a) and (b) show the room temperature Mössbauer spectra of BFN- KN solid solutions synthesized by the SSR and the SCR methods, respectively. These samples are the same samples as those of the samples used for collecting the XRD data, as shown previously in Fig. 1. The Mössbauer spectra of both the BFN samples, synthesized by the SSR and SCR routes show similar paramagnetic quadrupole doublet patterns, indicating that both the samples are similar. However, for the SSR synthesized solid solutions, along with the quadrupole doublet, we observed a Zeeman split sextet pattern. The Mössbauer spectra were fitted accordingly, and the obtained Mössbauer parameters are listed in Table 1. The quadrupole doublet is assigned to the Fe atoms present at the octahedral position (B-site) of the pseudo-cubic perovskite structure, and it has no magnetic interaction with any of its neighbouring atoms as its neighbouring atoms are not magnetic. This is the origin of the paramagnetic doublet observed in all our samples. However, it is interesting to understand the origin of the sextet observed in our solid solutions synthesized by the SSR route. Clearly, the intensity of this Zeeman sextet gradually increases as the intended BFN content in the solid solution decreases. This suggests that as the amount of intended BFN content decreases, the pseudo-cubic phase formation of the solid solution is not favouring. Furthermore, the hyperfine parameters obtained for this sextet (Table 1) matches with that of the α-Fe$_2$O$_3$ [53] [53–56], confirming that α-Fe$_2$O$_3$ phase is present in this SSR synthesized solid solution samples. Hence, it seems that the amount of the α-Fe$_2$O$_3$ phase is higher for the samples with lower intended content of BFN; this can be conceived from the fact that as the formation of pseudo-cubic phase is not favoured, higher amounts of the α-Fe$_2$O$_3$ precursor remains un-dissolved in the solid solutions. Note that this unreacted α-Fe$_2$O$_3$ phase is undesirable, and this is not observed in the XRD pattern. This suggests that these α-Fe$_2$O$_3$ phases are of sub-nanometric size. This observation clarified that although there are many reports show the formation of solid solutions via SSR route just by looking at their XRD patterns, and it does not give any clear picture of the sample.

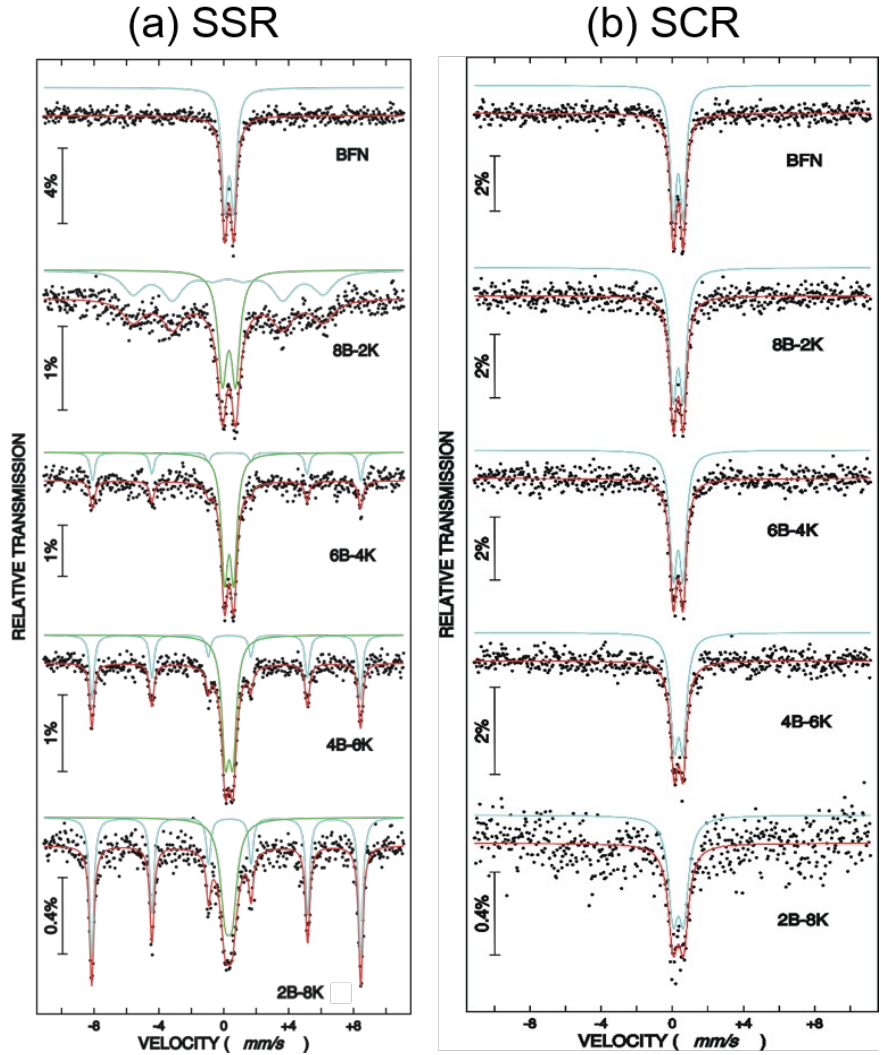

Fig. 2. Mössbauer spectra of the (x) BFN - (1-x) KN (x = 0, 0.2, 0.4, 0.6, 0.8) solid solutions, synthesized by (a) SSR route, and (b) SCR route.

Table 1: The hyperfine parameters obtained from the least-squares fitting of the Mössbauer spectra of the studied sample synthesized by the SSR route. The isomer shift (IS), quadrupole splitting or quadrupole nuclear level shift (QS), linewidth, magnetic hyperfine field ($B_{hf}$), spectral area (Area), and the 2$^{nd}$ to 3$^{rd}$ line intensity ratio are tabulated below. The estimated error bar in all the parameters is about 2% of the values obtained.

| Sample code | Sample name | Spectra | IS (mm/s) | QS (mm/s) | $B_{hf}$ (T) | Width (mm/s) | Area % | A23 |
|---|---|---|---|---|---|---|---|---|
| BFN | BFN | Doublet | 0.33 | 0.55 | | 0.44 | 100 | |
| 8B-2K | 0.8BFN-0.2KN | Sextet | 0.25 | 0.03 | 36.4 | 1.50 | 56 | 3.79 |
| | | Doublet | 0.32 | 0.81 | | 0.64 | 43 | |
| 6B-4K | 0.6BFN-0.4KN | Sextet | 0.26 | -0.18 | 51.2 | 0.32 | 25 | 2.26 |
| | | Doublet | 0.33 | 0.55 | | 0.49 | 74 | |
| 4B-6K | 0.4BFN-0.6KN | Sextet | 0.26 | -0.21 | 51.35 | 0.31 | 41 | 1.98 |
| | | Doublet | 0.33 | 0.46 | | 0.54 | 58 | |
| 2B-8K | 0.2BFN-0.8KN | Sextet | 0.26 | -0.22 | 51.4 | 0.32 | 61 | 2.1 |
| | | Doublet | 0.30 | 0.43 | | 0.73 | 39 | |

Table 2. The hyperfine parameters obtained from the least-squares fitting of the Mössbauer spectra of the studied sample synthesized by SCR route.

| Sample code | Sample name | Spectra | IS (mm/s) | QS (mm/s) | Width (mm/s) |
|---|---|---|---|---|---|
| BFN | BFN | doublet | 0.33 | 0.54 | 0.40 |
| 8B-2K | 0.8BFN-0.2KN | doublet | 0.33 | 0.55 | 0.47 |
| 6B-4K | 0.6BFN-0.4KN | doublet | 0.33 | 0.54 | 0.48 |
| 4B-6K | 0.4BFN-0.6KN | doublet | 0.35 | 0.48 | 0.51 |
| 2B-8K | 0.2BFN-0.8KN | doublet | 0.33 | 0.59 | 0.64 |

The Mössbauer spectra of all the solid solution samples prepared by the SCR route clearly show a doublet pattern each, indicating the formation of phase pure double perovskite compound for all solid solutions composition. The spectra are fitted by a quadrupole doublet and the corresponding hyperfine parameters obtained from the fittings are given in Table 2. Furthermore, this confirms that as the intended BFN content in the samples decreases, the required calcination temperature to obtain phase pure samples also decreases. These temperatures are also lower than that of the calcination temperatures of the corresponding samples prepared by the SSR method. Clearly, due to homogenous mixing of the Fe and also other ions during the solution stage, the cations go to their respective positions, facilitating ease formation of the double perovskite structure of the BFN-KN solid solution at low temperature, without any impurity phase. This includes the uniform distribution of Fe atoms in the solid solution giving rise to the paramagnetic phase.

From the XRD and Mössbauer analysis, it has been confirmed that BFN-KN solid solution samples synthesized by SCR are phase pure. Due to the molecular level mixing of precursors in the SCR, phase pure samples are formed at reduced calcination temperature. Such mixing is not possible in the conventional SSR. From our study, for BFN-KN solid solution samples, SCR is more effective than the SSR. The phase pure samples obtained by SCR were further investigated for structural, microstructural, optical, and dielectric properties.

## 3.3 Raman spectroscopy

To understand local lattice distortion and long-range ferroelectric order in our solid solutions samples, Raman spectroscopic measurements were performed [57, 58]. Fig. 3 shows the normalized Raman spectra of our (x)BFN-(1-x)KN(x = 0, 0.2, 0.4, 0.6, 0.8, 1) solid solution samples. The Raman active modes for BFN and KN are identified as shown in the Fig. 3 and listed in Table 3.

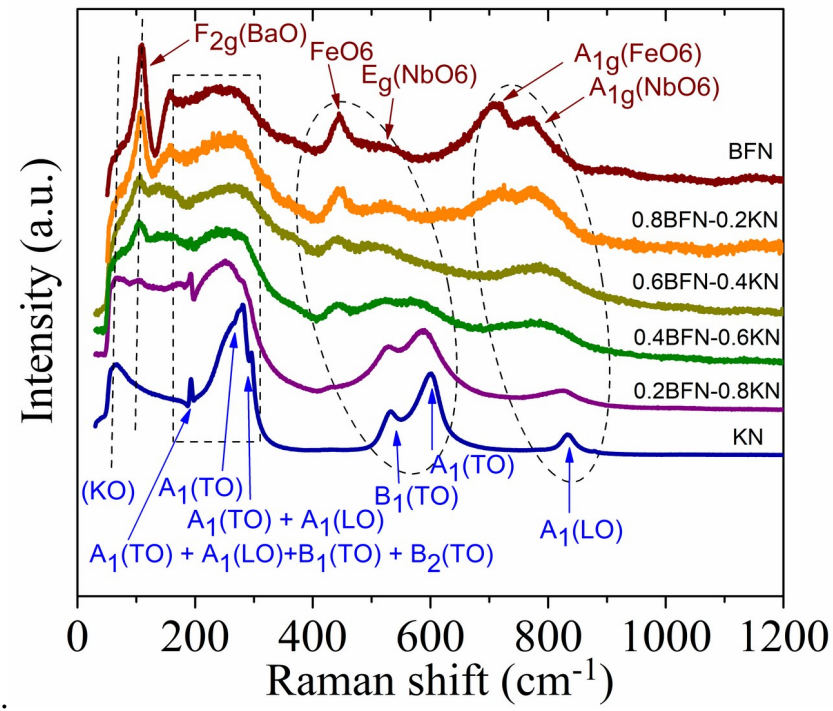

Fig. 3. The measured Raman spectra for the (x)BFN - (1-x)KN (x = 0, 0.2, 0.4, 0.6, 0.8, 1) solid solution samples synthesized by SCR method.

According to group theory analysis, for our $KNbO_3$ sample, there are 12 optical modes expected for the mm2 point group: $4A_1 + 4B_1 + 3B_2 + A_2$, which can be separated into translational modes of isolated $K^+$ and internal modes of the $NbO_6$ octahedra[59–62]. The mode $A_1(LO)$ observed at 833.29 cm$^{-1}$ is the symmetric breathing mode of the $NbO_6$ octahedron. In this vibrational mode, the Nb-O in the octahedral oscillate in the way as given in Fig. 4(I): (a) the breathing out of O octahedron, (b) the equilibrium position of O octahedron, and (c) the breathing in of O octahedron. Modes $A_1(TO)$ and $B_1(TO)$ observed at

Table 3. The list containing the assignment of the Raman mode observed in our studied samples. The modes are separately observed for KN [59–62] and BFN [63–67]. The symbols used are, TO: Transverse Optic mode, LO: Longitudinal Optic mode.

| Raman Shift (cm$^{-1}$) | Raman Modes | Atoms involved | Description |
|---|---|---|---|
| **KNbO$_3$ (KN)** | | | |
| 64 | B$_2$(TO) | K-O octahedral | K with respect to O octahedral |
| 193 | A$_1$(TO)+A$_1$(LO)+ B$_1$(TO)+ B$_2$(TO) | Mixed mode | Long-range ordering |
| 295 | A$_1$(LO) + A$_1$(TO) | Mixed mode | Long-range ordering |
| 281 | A$_1$(TO) | O-Nb-O Bending | NbO$_6$ Bending |
| 532 | B$_1$(TO) | O-Nb-O Stretching mode | NbO$_6$ Stretching. If 4 coplanar O atoms closer, then 2 perpendicular atoms go away. |
| 601 | A$_1$(TO) | O-Nb-O Stretching mode | NbO$_6$ Stretching. If 4 coplanar O atoms come closer, then 2 perpendicular atoms go away. |
| 833 | A$_1$(LO) | NbO$_6$ | Symmetric breathing mode |
| **BaFe$_{0.5}$Nb$_{0.5}$O$_3$ (BFN)** | | | |
| 109 | F$_{2g}$ | Ba-O | Ba with respect to O octahedral |
| 170 to 330 | | Fe and Nb | Cations off-center shift |
| 443 | | O-Fe-O Stretching mode | FeO$_6$ Stretching mode |
| 520 | E$_g$ | NbO$_6$ Stretching mode | NbO$_6$ Stretching mode |
| 710 | A$_{1g}$ | FeO$_6$ | Symmetric breathing mode |
| 770 | A$_{1g}$ | NbO$_6$ | Symmetric breathing mode |

601 cm$^{-1}$ and 532 cm$^{-1}$ are the stretching modes of the NbO$_6$ octahedron. In this mode, if four coplanar O atoms come closer, then the remaining two O atoms move further, as shown in Fig. 4 (II). The intensity of 601 cm$^{-1}$ modes is more as compare to 532 cm$^{-1}$. This is due to the non-isotropic nature of coplanar O atoms in NbO$_6$ octahedral of orthorhombic KNbO$_3$. The sharp modes observed at 193 cm$^{-1}$ and 295 cm$^{-1}$ are mixed modes corresponds to the A$_1$(TO)

+$A_1$(LO)+ $B_1$(TO)+$B_2$(TO) and $A_1$(LO)+$A_1$(TO), respectively. These occur due to the long-range ordering of dipoles (displaced Nb with respect to O octahedron form dipoles) and characteristics of $KNbO_3$ ferroelectric nature. Mode $A_1$(TO) at 281 cm$^{-1}$ and $B_2$(TO) at 64 cm$^{-1}$ correspond to the $NbO_6$ octahedron bending and stretching of $K^{1+}$ ions with respect to O octahedron, respectively.

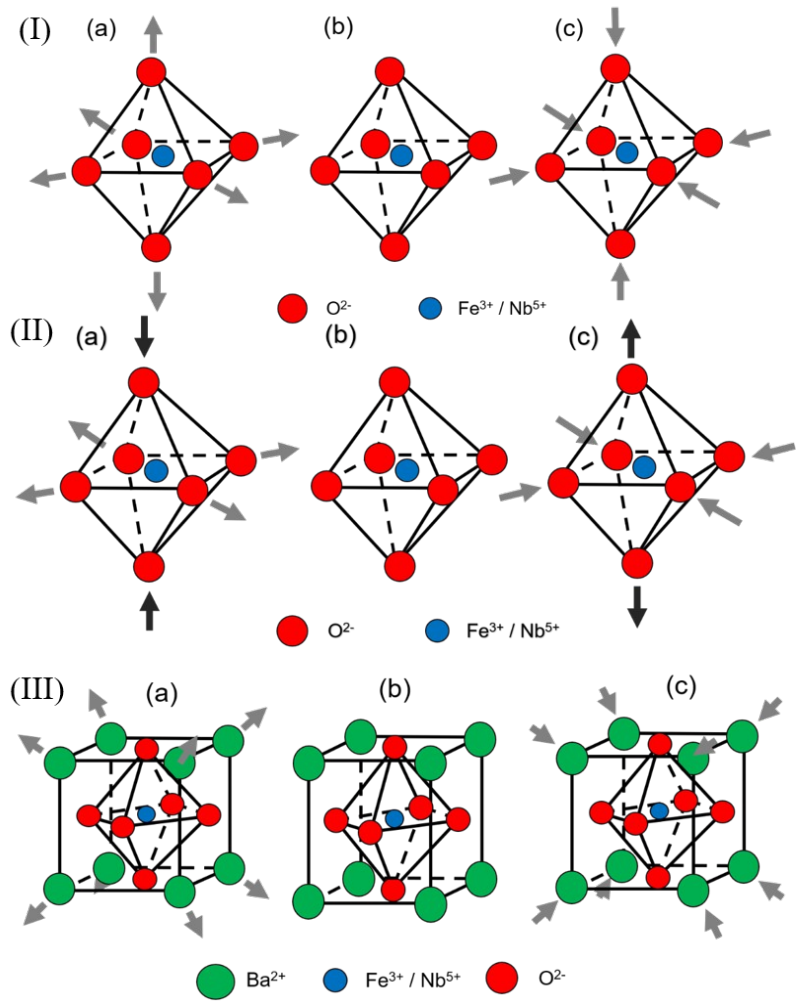

Fig 4: (I) symmetric breathing mode of Oxygen octahedron, (II) Stretching mode of Oxygen octahedron, and (III) the motion of $Ba^{2+}$ ions against the Oxygen octahedron.

For our $BaFe_{0.5}Nb_{0.5}O_3$ sample, point group of m-3m, there are 9 Raman active mode $2F_{2g} + E_g + A_{1g}$ [63–67]. The mode $F_{2g}$ at 109 cm$^{-1}$ corresponds to the vibration of $Ba^{2+}$ with respect to the oxygen octahedra as shown in Fig. 4 (III). Fig. 4 (III) (b) shows the equilibrium position, (a) $Ba^{2+}$ moves away from the octahedron and (c) towards the octahedron. The modes that occur between 170 and 330 cm$^{-1}$ correspond to the Fe and Nb cation off-center shift with respect to the octahedron. The sharp mode at 443 cm$^{-1}$ and doubly degenerate broad mode $E_g$ at 520 cm$^{-1}$ correspond to the FeO$_6$ and NbO$_6$ octahedral stretching modes, respectively. Fig. 4(II) shows such octahedral stretching mode for FeO$_6$ and NbO$_6$. The mode $A_{1g}$ occurs at 710 cm$^{-1,}$ and 770 cm$^{-1}$ are symmetric breathing modes of FeO$_6$ and NbO$_6,$ respectively, as shown in Fig. 4(I).

In our (x) BFN - (1-x) KN (x = 0, 0.2, 0.4, 0.6, 0.8, 1) solid solution, it observed that Raman spectra of solid solutions shows changes in the modes of KN as the concentration of BFN increases (Fig. 3). The $B_2$(TO) mode at 64 cm$^{-1}$ starts disappearing, and $F_{2g}$ sharp mode at 109 cm$^{-1}$ starts appearing with increased BFN concentration, this due to the increasing Ba$^{2+}$ at the K$^{1+}$ site. The sharp (mixed) modes observed at 193 cm$^{-1}$ and 295 cm$^{-1}$ start disappearing, and it forms the broad peaks due to change in long-range ordering into the disordered arrangement of cations. This indicates the transition of KN from displacive ferroelectric to order-disorder ferroelectric. As BFN concentration increases, two stretching modes (532 cm$^{-1}$ and 601 cm$^{-1}$) of NbO$_6$ merged, and a single broad peak is observed at 520 cm$^{-1}$; this due to the transition of non-isotropic NbO$_6$ octahedral to the isotropic octahedron. This further confirmed the phase transition from orthorhombic to the pseudo-cubic structure. The appearance of sharp mode at 443 cm$^{-1}$ for FeO$_6$ stretching shows that as Fe$^{3+}$ goes at the Nb$^{5+}$ site as BFN concentration increases in KN. This further confirmed by splitting of symmetric breathing mode of NbO$_6$ octahedron (833 cm$^{-1}$) into two symmetric breathing modes of FeO$_6$ and NbO$_6$ (710 cm$^{-1}$ and 770 cm$^{-1,}$ respectively). As BFN concentration increases in KN, breathing modes start shifting towards lower energy, indicating mode softening. To conclude from Raman analysis, BFN increases disorder arrangement of Nb site by increasing Fe concentration in KN. This disturbs the long-range dipolar ordering of KN and converts it into the disordered arrangement, making the displacive ferroelectric KN into an order-disorder ferroelectric in nature. The stretching modes show that BFN converts orthorhombic KN into the pseudo-cubic structure. Furthermore, it confirmed that as BFN concentration increases in the KN, Ba$^{2+}$ goes to the K$^{1+}$ site and Fe$^{3+}$ goes to the Nb$^{5+}$ site and forms phase pure solid solution samples.

## 3.4 FTIR spectroscopy

In the Infrared (IR) spectroscopy, through the interaction between dipoles (molecules) and IR radiation, the vibrations (stretching and bending) of bonded atoms within the sample were determined. The vibrations of cations that are bonded to the oxygen ions in perovskites produce different vibrational frequencies that appear as IR absorption bands in the FTIR spectrum. The spectral bonds' position depends on the strength and type of bonding, coordination between the ions involved, the crystal/local symmetry, and type (valence and mass) of the ions involved. The principle of FTIR spectroscopy can be well understood based on the spring-model of vibration of atoms, wherein the wave number $v$ of vibration is defined as [68]:

$$v = \frac{1}{2\pi c}\sqrt{k/\mu} \quad (1)$$

where $k$ is the spring (force) constant, $m$ is the reduced mass of ions involved, and $c$ is the velocity of light. The force constant is a measure of bond strength. The elastic and thermal properties of the samples depend directly on the bond strength.

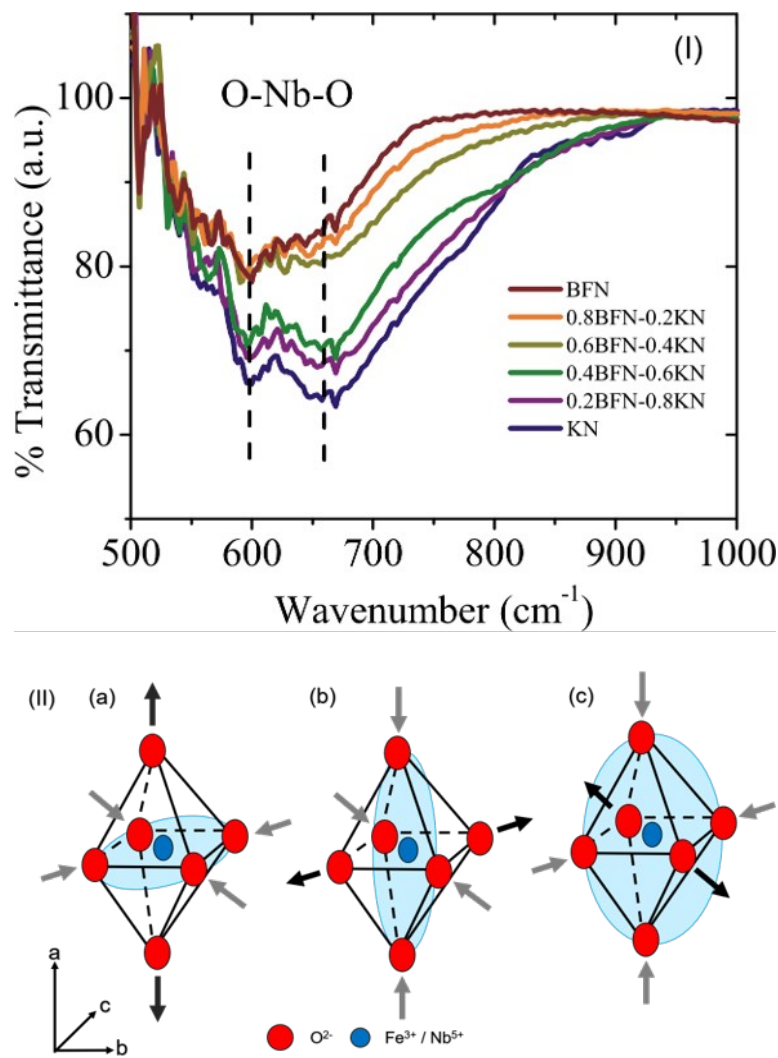

Fig. 5. (I) The FTIR spectra of (x) BFN - (1-x) KN (x = 0, 0.2, 0.4, 0.6, 0.8, 1) solid solutions (II) O-Nb-O (and O-Fe-O) stretching modes: (a) stretching along 'a' direction and contraction in the cb plane, (b) stretching along 'b' direction and contraction in the ca plane, and (c) stretching along 'c' direction and contraction in the ab plane, in $NbO_6$ (and $FeO_6$) octahedron

of KN (and BFN). The light blue colored plane shows the contraction of the 4 coplanar O atoms.

The FTIR spectra of our (1-x) BFN-(x) KN (x = 0, 0.2, 0.4, 0.6, 0.8, 1) solid solutions measured between the wavenumber from 500 to 1000 cm$^{-1}$, is shown in Fig. 5(I). For the KN sample, two transmittance peaks positioned at 600 cm$^{-1}$ and 655 cm$^{-1}$ are observed, but only one peak is observed for the BFN sample (at 600 cm$^{-1}$). These peaks correspond to the O–Nb–O stretching mode of NbO$_6$ octahedron of KN [69–71], (this also includes the O-Fe-O stretching modes in BFN). As KNbO$_3$ belongs to the *Amm2* space group with orthorhombic crystal structure, the lattice constants are related as a ≠ b ≈ c [72]. The schematic of O–Nb–O stretching mode of octahedral corresponds to 600 cm$^{-1}$ peaks is shown in Fig. 5(II)(a) and for the peak at 655 cm$^{-1}$ is shown in Fig. 5(II)(b) and (c). As BFN concentration in KN increases, the peak at 655 cm$^{-1}$ starts disappearing; this indicates the change in the lattice constant values. Therefore, this observation shows that BFN converts orthorhombic KN in the pseudo-cubic structured solid solution. From previously discussed Raman results and FTIR results show that, as BFN concentration increases, more $Fe^{3+}$ occupies the $Nb^{5+}$ site. This disturbed the long-range ordering of dipoles created by the $Nb^{5+}$ cations in the octahedron formed by the $O^{2-}$ anions, making displacive ferroelectric KN into the disordered one. The crystal structure of KN changes from orthorhombic to pseudo-cubic too. Overall, due to the formation of solid solution, there is a tilting and/or an expansion of the $O^{2-}$ octahedra resulting from the partial occupation of bigger $Ba^{2+}$ cations at the A-site of the perovskite structure.

## 3.5 Optical properties

The UV-Vis reflectance spectra for the (x) BFN - (1-x) KN(x = 0, 0.2, 0.4, 0.6, 0.8, 1) solid solutions is shown in the Fig. 6(a). The reflectance data were used for obtaining the Kubelka-Munk function [73–75], as shown Fig. 6(b), which is used for obtaining the bandgap for our solid solutions. The variation of the energy band-gap values of our BFN-KN solid solutions is shown in Fig. 6(c). It is observed that, as the concentration of BFN increases, the energy band gap gradually decreases from 3.34 eV for KN to 1.88 eV for the BFN sample. In perovskite structured materials, the structural changes such as tilting, contraction or expansion of the oxygen octahedra, lattice strain, etc. are the controlling parameters for the modification in the energy bandgap values [76–79]. From the XRD, Raman, and FTIR analysis, we observed that as the BFN concentration 'x' increases, the lattice distortion is observed for KN and vice versa. This lattice distortion through octahedral tilting, which originates due to the occupation of bigger $Ba^{2+}$ cations at the A sites, seems to be the origin of the decrease in energy bandgap in our studied solid solution samples[60]. As explained in the XRD results, we observed an increase in lattice parameters as BFN content in the solid solution increases, which is in line with earlier reports[60].

Note that in the Kubelka-Munk plots for all our samples (Fig. 6 (b)), we have clearly observed three important features. As we go from the lower energy towards higher energy values, we observe: (1) a tail at the lower energies, which diminishes as the energy decreases further, (2) a sharper rise in the medium energies, and (3) a linear rise at the higher energies before reaching the saturation. The low energy tail is known as the Urbach tail, which originates due to the presence of dense defects, especially in the grain boundary regions [80–82]. The sharper rise in the medium energy range is due to the modification of the band gap closer to the band edge region, which arises due to the tilting of the oxygen octahedra in perovskites and variation in lattice parameters (or expansion and contraction of the octahedra) due to occupation of bigger/smaller cations at the A- or B-sites [60]. Furthermore, the faster or slower absorption in the linear region depicted as a higher or lower slope of the linear region can be assigned to the lattice distortion occurring due to variation in the lattice strain in the samples.

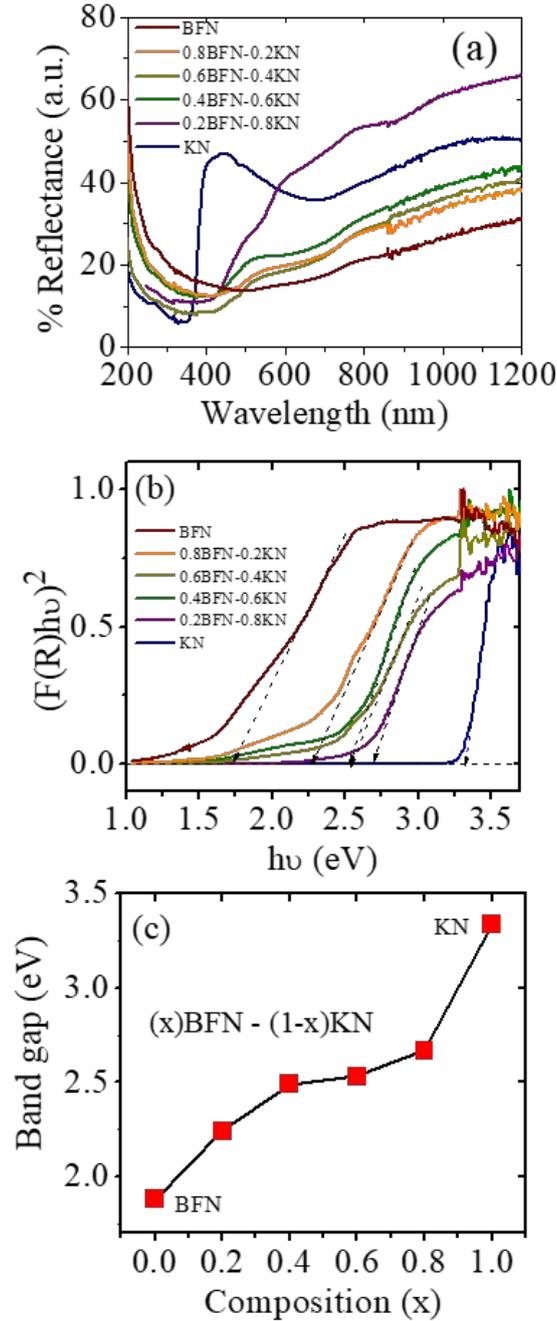

Fig. 6(a) UV-Vis reflectance spectra, (b) plot of the Kubelka-Munk function in the Tauc formalism for obtaining the bad gap, and (c) obtained energy band gap values for our (x)BFN - (1-x)KN (x = 0, 0.2, 0.4, 0.6, 0.8, 1) solid solutions.

Comparing the features observed in different samples, we can rationalize that the slope of the high linear region decreases as the BFN content 'x' increases. This feature has a direct correlation with the variation of linewidth of the XRD peaks (see, e.g., Fig. 1(c)). Hence, this decrease in the slope of the linear region in Fig. 6(b) is originating due to the variation in lattice strain in the samples (due to the occupation of bigger $Ba^{2+}$ cations).

Comparing the sharper rising middle region of the plot for all our samples, it seems that this reason is associated with the variation in lattice distortion, tilting of the octahedron, and expansion/contraction of the oxygen octahedron as a result of variation in chemical composition in the solid solutions. This may be due to the sharing of the A-sites by $Ba^{2+}$ and $K^+$ cations instead of only $Ba^{2+}$ or $K^+$ cations, and the B-sites by the $Fe^{3+}$ and $Nb^{5+}$ cations instead of only $Nb^{5+}$ in either $KNbO_3$ or $BaFe_{0.5}Nb_{0.5}O_3$ [76–79]. Similarly, a comparison of the Urbach tail regions suggests that there could be higher disorder in the solid solutions (not in the single-phase samples, KN/BFN); these disorders are generally present in the grain boundary / amorphous regions of the samples. This behavior can be rationalized by considering the SEM images, as will be discussed in a later section. It should be noted in all samples, all these three effects are present and differently contributing to the optical absorption in different samples.

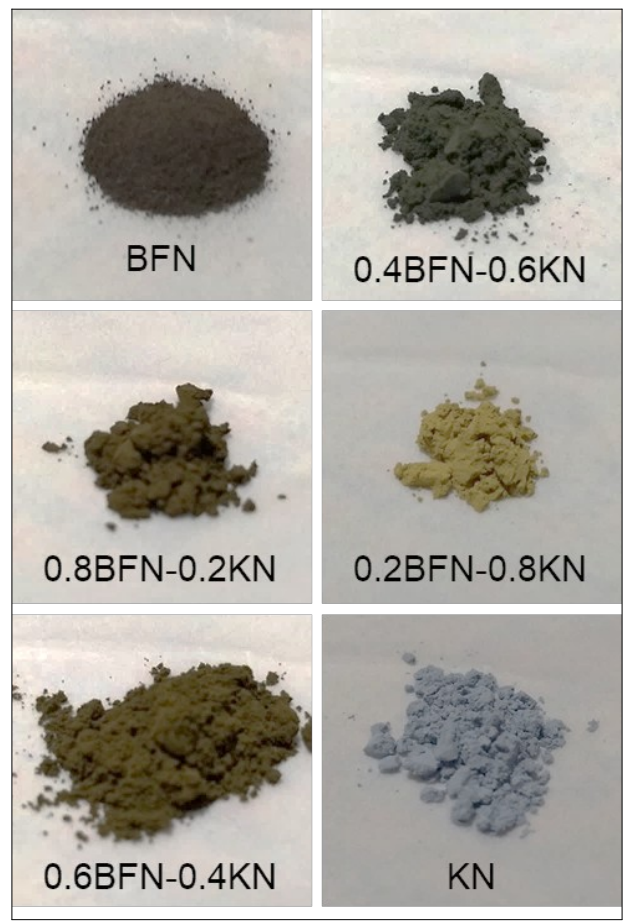

Fig. 7. The optical picture of the synthesized (x) BFN - (1-x) KN (x = 0, 0.2, 0.4, 0.6, 0.8, 1) samples showing the variation in colour of the solid solution samples as the composition 'x' changes.

Fig. 7 shows the optical image of the (x)BFN - (1-x)KN (x = 0, 0.2, 0.4, 0.6, 0.8, 1) solid solutions. For KN (x=0) powder, light blue color is observed. As the BFN concentration increases, the color of solid solution powder changes as follows: yellow for x = 0.2, dark green for x = 0.4, light shades of brown for x = 0.6 and x = 0.8 and finally blackish-brown for x = 1 (BFN). This color change can be majorly due to the changes in the lattice (expansion of the lattice/tilting of oxygen octahedra) due to gradually higher occupation of the bigger $Ba^{2+}$ cation at the A-sites and an increased concentration of transition metal ion $Fe^{3+}$ at the B-sites. This results in the observed changes in the energy bandgap values from 3.34 eV to 1.88 eV and correspondingly a change in the color of the samples with increased BFN concentration.

### 3.6 Morphology and density of sintered pellets

In the study of the dielectric properties of the samples, a dense and phase pure pellet is required. Hence, after obtaining phase pure powder samples through calcination, the powder samples are pressed into pellets and sintered at an appropriate temperature to obtain a dense and phase pure pellet. This pellet is then characterized to measure the density of the pellet and also study the dielectric behavior of this pellet so that the Maxwell-Wagner (M-W) polarization is reduced and the proper dielectric polarization behavior of the material can be understood. As the M-W polarization is originated mostly due to the structural defects in the grain boundary regions where these defects and voids lead to leakage current, by sintering the pellets, sharp grain boundaries can be obtained, which can reduce the M-W polarization. Further discussion on the dielectric behavior of our samples will be discussed in the next section. Here, the SEM micrographs of the pellets of the BFN-KN solid solutions sintered at different temperatures, as indicated, are shown in Fig. 8. Clearly, the SEM micrograph of the pure BFN pellet sintered at a high temperature of 1250 °C shows well-connected grains with sharp grain boundaries making the pellet dense. The SEM image of the pellets of the solid solution samples shows diffused regions between the particles, suggesting that the grain boundary region is not sharp and it contains voids/disordered phases at the grain boundaries. For pure KN, the SEM image of the pellet shows no diffused region, although the sample is phase pure and having sharp Bragg peaks. This may be due to the fact that although the temperature used for sintering this KN pellet (850 °C) is perfect for the formation of pure KN phase and for avoiding the formation of any impurity phases at higher temperatures, this temperature is not sufficient for densification of the sample through thermal diffusion.

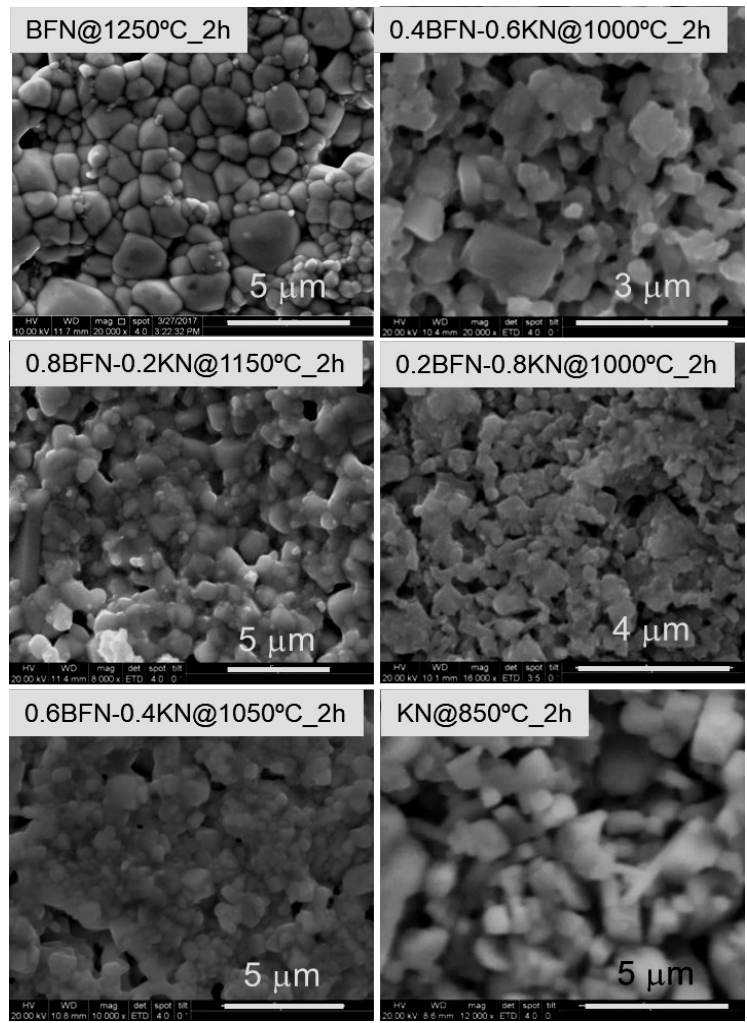

Fig. 8 SEM micrograph of the pellets of (x) BFN - (1-x) KN(x = 0, 0.2, 0.4, 0.6, 0.8, 1) solid solutions after sintering at suitable temperatures (in ºC), as indicated.

As density is another important factor controlling the dielectric properties of a sample, we have measured the density of the sintered pellets by Archimedes principle, and the results are shown in Fig. S3. It is observed that the relative density (as compared to the theoretical density) of the KN (x = 0) pellet is ~ 89 %, 92 % for x = 0.2, 91% for x= 0.4, 92% for x = 0.6, 95 % for x = 0.8 and 96% for x=1 (BFN). It seems that the increase in density observed for the pellets with increasing x, is majorly due to reduced numbers of pores because, with higher x, the samples are sintered at a higher temperature which facilitates thermal diffusion in the formation of a dense pellet. Explicitly, the KN pellet has a large number of pores, whereas BFN has fewer numbers of pores; for intermediate solid solutions, pores decrease with higher sintering temperature and the amount of BFN phase.

## 3.7 Dielectric properties

To explore the dielectric behavior of a material, understanding the type of dielectric polarization and its frequency dependence is very important. Hence, to investigate the dielectric constant and to find out the type of dielectric polarization that occurs in our (x)BFN - (1-x)KN (x = 0, 0.2, 0.4, 0.6, 0.8, 1) solid solutions, we have measured the frequency-dependent dielectric constant at RT and analyzed those through different formalism as it will be discussed below. Note that the samples used here are the (x)BFN - (1-x)KN (x = 0, 0.2, 0.4, 0.6, 0.8, 1) solid solutions synthesized via SCR route. Fig. 9(a, b) shows the real part ($\varepsilon'(\omega)$) and the imaginary part ($\varepsilon''(\omega)$) of the dielectric constant of all our samples, respectively. The corresponding Cole-Cole plots for all the solid solutions are shown in Fig. 9 (c).

For KN (x = 0) ceramic sample, the $\varepsilon'$ value was ~ $5 \times 10^3$ at 40 Hz frequency, but this value decreases faster (as the frequency increases) than the other samples and reaches a value of ~ 70 at $10^8$ Hz frequency (Fig. 9(a)). The origin of the low-frequency behavior of the $\varepsilon'$ can be assigned to the Maxwell-Wagner (M-W) type of polarization mechanism (along with small contributions from the dipolar and electronic polarization) [39, 43, 83–86]. At higher frequencies, where it reaches almost a constant value, it remains to be a combination of dipolar (as KN is a classical ferroelectric material) and electronic polarization, although a contribution from the tail of the M-W polarization variation is also present. For the solid solution sample with x =0.2, the $\varepsilon'$ values are higher at all the measured frequencies than that of the KN sample, and the rate of decrease in $\varepsilon'$ value with increasing frequency is much slower. Interestingly, for this sample, the $\varepsilon'$ value observed at the highest measured frequency region is the highest among all our samples. Understanding the origin of this behavior is important for understanding the nature of polarization occurring in our materials, which will be discussed in the paragraph below. The variation of the $\varepsilon'$ values observed for all other samples (with x = 0.4, 0.6, 0.8, and 1) are similar in that they consist of three particular features in three different frequency regions: a low very low-frequency dispersion (< ~ $10^3$ Hz), followed by a step and dispersion in the medium frequency region (between ~ $10^3$ Hz and ~ $10^7$ Hz) before reaching at a low but almost constant $\varepsilon'$ value at very high frequencies ($> \sim 10^7$ Hz) as the third feature. Herein, we explore the origin of this polarization behavior in our samples. The very low-frequency dielectric dispersion for the solid solution samples with

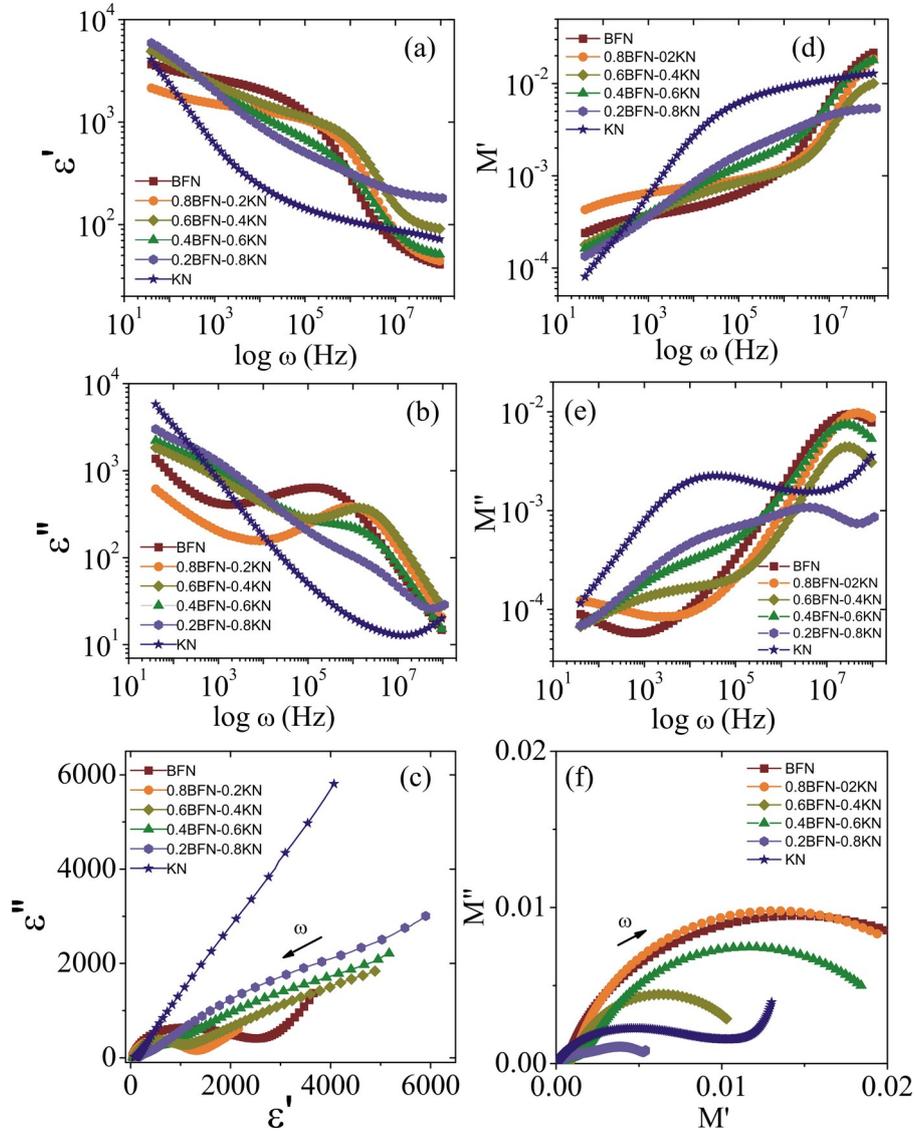

Fig. 9(a) The real part of dielectric constant ($\varepsilon'(\omega)$), (b) Imaginary part of the dielectric constant ($\varepsilon''(\omega)$), (c) the Cole-Cole plot for the dielectric constants, (d) real part of dielectric modulus ($M'(\omega)$), (e) Imaginary part of the dielectric modulus ($M''(\omega)$), and (f) the Cole-Cole plots for the dielectric modulus of the pellets of the (x)BFN - (1-x)KN (x = 0, 0.2, 0.4, 0.6, 0.8, 1) solid solutions measured at RT. The arrow in (c) and (f) indicate the direction of increasing frequency.

x= 0.4 to 1.0 can be assigned to the Maxwell-Wagner polarization arising due to the amorphous-like or densely defected regions of the sample, mostly the grain boundaries. This type of polarization occurs mostly due to the accumulation of charge carriers at the interface between grain and grain boundaries in the sample or the charge accumulation at the surface of the grains/particles. Note that we assumed the grain boundary to have a finite width. As more defects are present at the grain boundaries and on the surfaces, the resistance to diffusion of these charged species is high (and the time taken for the polarization/charge accumulation is high), and the corresponding polarization manifests only at the lower frequencies of the applied ac field. After the careful observation of these high-frequency values, we can rationalize that the $\varepsilon'$ value is highest for the solid solution with x= 0.2, but the $\varepsilon'$ value gradually decreases on both sides of this composition, i.e., as the content of BFN either decreases or increases.

It is clear that the low-frequency dispersion of the dielectric constant originates due to the Maxwell-Wagner type polarization associated with the charge accumulation at the grain boundaries. Let us now discuss the behavior of $\varepsilon'(\omega)$ observed in the medium frequency region (in Fig. 9(a)), where we have seen a step followed by a frequency dispersion in the dielectric constant. This observed feature suggests that the polarization can have its origin in the dipolar polarization. However, the measured value of $\varepsilon'$ observed for our classical ferroelectric sample (KN sample, x = 0) is much lower than all other samples in the same frequency region. This is a contradiction because KN is a classical ferroelectric material at RT. The question now arises: wherefrom this high dipolar polarization (or high value of $\varepsilon'$) could originate in the samples having a certain amount of BFN? As discussed previously, all the solid solution samples, including BFN, show pseudo-cubic structure at RT. Hence, we expect the polarization for these samples (having pseudo-cubic structure) should be low as compared to that of the KN sample. Furthermore, if the observed polarization is dipolar in nature, then its existence in pure BFN sample should have been observed up to a higher frequency value than that for all the solid solution samples, i.e., this step would have been extended up to higher frequencies for the samples with higher BFN content. However, we have not observed this. Hence, the origin of this step-like behavior for the BFN sample and for all the solid solution samples is not due to dipolar polarization.

To understand the type of polarization associated with this step like $\varepsilon'(\omega)$ behavior, let's consider, firstly, the high-frequency behavior of $\varepsilon'(\omega)$ observed in all our samples. As stated earlier, the value of $\varepsilon'(\omega)$ observed for our solid solution with x = 0.2 is the highest. On both sides of this composition, the dielectric constant (at the highest frequency region) decreases (Fig. 9(a)). In analogy with our previous discussion, we can also confirm that the $\varepsilon'$ (ω) observed for this high-frequency region is also not associated with any dipolar polarization. Had it been dipolar polarization, the absolute value of ε′(ω) should have been the highest either for the pure KN or pure BFN sample. However, the higher dielectric constant (polarization) observed for the KN sample than that of the BFN sample suggests that the higher dielectric constant for KN originates due to the dipolar nature of the polarization for this sample, but it is not the case for BFN. Here, let us consider the results obtained from the characterizations of the samples. From the XRD analysis, as we have mentioned, the XRD line-width (FWHM) of the samples (Fig. 1(c)) exhibits a very similar trend as that of the high-frequency dielectric constant (Fig. 9(a)). The XRD line-widths are related to the strain variation in our samples, which leads to the modification in the local electronic properties and hence, the optical properties [62]. This change in optical properties is manifested as the variation in the slope of the linear region of the Kubelka-Munk plot, as discussed previously (Fig. 6(b)). Furthermore, it can be rationalized that the electronic property variation due to local strain in the samples won't change the dielectric behavior of the whole sample, but rather the frequency behavior of the dielectric constant would be closer to that of a perfect sample (with no local strain). Hence, we can assign this frequency dependence to the Maxwell-Wagner type polarization occurring locally in the strained regions of the samples. Note that this local strain is present well within the grains (i.e., the strain at the grain boundaries is not necessarily involved here).

In view of the above discussion, we can now rationalize that the 'step followed by the frequency dispersion' observed in the medium frequency region (in the samples with x ≥ 0.2) is not due to dipolar polarization occurring in the samples, but it has its origin in the change in electronic properties of the samples due to occupation of the larger $Ba^{2+}$ cations (along with $K^+$ cations) at the A-sites of the perovskite structure, and also the occupation of $Fe^{3+}$ cations at the B-sites (along with $Nb^{5+}$ cations). As we have observed in the XRD results that the Bragg peaks shift gradually to lower Bragg angles (Fig. 1), i.e., the lattice parameter increases as the BFN content 'x' in the solid solution increases. Furthermore, in the results of the FTIR spectral analysis (Fig. 5), we have observed the tilting and expansion of the $O^{2-}$ octahedra, as gradually more $Ba^{2+}$ cations occupy the A-site (with increasing x). Furthermore, in the Kubelka-Munk plot (Fig. 6(b)), we have observed a region between the Urbach tail and linear region, having a sharper rise in optical absorption. All these results establish that there is a modification in the electronic properties (modification of the band structure) as a result of the occupation of bigger $Ba^{2+}$ cations at the A-sites. This modification in the electronic band structure due to the local variation in lattice parameter and/or tilting and expansion of the $O^{2-}$ octahedra is the origin of the observed step followed by the dispersion in the dielectric constant observed in our solid solution samples (Fig. 9(a)). Note here that this local octahedron-tilting and octahedron-expansion are not necessarily coupled with the local strain, i.e., they can be energetically different and have different frequency responses than the charge carriers originating due to the strain in the samples. Hence, the charge carriers originating due to these kinds of defects are having different activation energy, and hence they respond differently at different frequencies of the applied ac field. Hence, this type of polarization behavior can still be considered as the Maxwell-Wagner type of polarization. Note that the expansion/tilting of octahedra occurs inside the grains of the sample (i.e., the octahedral modification at the grain boundary regions is not necessarily involved here).

Taking all the results together, we can now understand that the high value of dielectric constant (or polarization) at the higher frequency region for the sample with x=2 is due to the response of the charge carriers associated with the large strain present in the sample (as seen from the FWHM of the XRD peaks for this sample). Furthermore, the overall nature of this dielectric response is close to that of the KN sample, i.e., the step-shaped frequency response of the dielectric constant is not pronounced. This is due to the fact that the lower amount of doping is not sufficient to increase the number of octahedra that are tilted/expanded. Furthermore, although the pure BFN sample is well crystalline, the alternate presence of the $Fe^{3+}$ and $Nb^{5+}$ cations at the octahedral (B-) sites makes the chemical inhomogeneity of the structure, along with the associated octahedral modification. Hence, this sample also shows the step-like behavior of the dielectric constant in the middle-frequency range. As mentioned earlier, this step-like frequency behavior of the dielectric constant for the pure BFN and/or the associated solid solutions creates ambiguity on its origin and the associated polarization. It is worth mentioning here that the local motion of the charged species responsible for this polarization functions exactly similar to that of the dipoles in the unit cell of the ferroelectric materials. Hence, in many such pseudo-cubic materials, the origin of this step-like frequency behavior of the dielectric constant is mistakenly assumed as dipolar in nature.

To establish the above concepts further, in Fig. 9(b), we have plotted the corresponding imaginary part of the dielectric constant versus frequency. All the observed behavior, including a relaxation peak observed at about a few MHz frequency range, is seen as per the expectation. Furthermore, the relaxation peak shifts towards a lower frequency as the BFN content in the solid solution increases. The appearance of the relaxation peak suggests that the carriers are confined/bound to the local regions within the grains where the octahedra tilting/expansion is present. The Cole-Cole plot is shown in Fig. 9(c). The Cole-Cole plot for our KN sample shows almost a sharply raising low-frequency tail (almost a straight line), i.e., it has a maximum contribution from the interfacial polarization. This behavior can be explained by a direct correlation with the SEM micrographs (Fig. 8), that the particles are not well-connected with each other, but they are separated from each other by a void; as a result, the charge carriers are not allowed to move away from the surface of the particles through the voids. Hence, they show very high capacitance (imaginary part), and also, the associated dc resistance to motion of these charged species is very high. Note that the dc resistance is the value on the real permittivity axis where the observed incomplete low-frequency semicircle intersects. This aspect is more clear in the impedance plots as shown in Fig. S4. For the other samples, as they show denser microstructure and the amount of voids in the sample decrease with an increase in BFN content, the surface capacitance, and the corresponding dc resistance are observed to gradually decrease. By careful observation of the Cole-Cole plot, we can clearly observe that apart from this low-frequency incomplete semicircle, there are two more semicircles at higher frequency regions. This is very clear from the Cole-Cole plot of the sample with x=0.2. The smaller semicircle observed in the mid-frequency region is associated with the octahedra tilting/expansion, and the smallest semicircle seen in the very high-frequency region is associated with the local strain in the samples, as discussed earlier. Furthermore, to enhance the visibility of the hidden features (especially at the high-frequency region), the corresponding modulus plots are shown in Fig. 9(d, e, f). Clearly, the information we obtain from the modulus plots clarifies our explanations and is well in agreement with all our results.

## 4. Conclusion

In this study we characterized the $(x)BaFe_{0.5}Nb_{0.5}O_3$ - $(1-x)KNbO_3$ (x = 0, 0.2, 0.4, 0.6, 0.8, 1) (BFN-KN) solid solutions prepared by solution combustion (SCR) route and showed its superiority in synthesizing phase pure ceramic samples and solid solutions than the synthesis by using the solid-state reaction (SSR) route. The BFN-KN solid solutions synthesized by the SCR route are phase pure due to molecular-level mixing of precursors than the samples prepared through the SSR route. The XRD pattern reveals that even a small amount of BFN addition changes the structure from the orthorhombic for KN towards the pseudo-cubic structure of BFN, with a gradual increase in the lattice parameter. Interestingly, the Mössbauer spectra for the SSR synthesized samples show a paramagnetic doublet along with a small contribution from a sextet, which corresponds to the unreacted $\alpha$-$Fe_2O_3$ impurity phase, whereas a single quadrupole doublet is observed for the samples synthesized by the SCR route. As Fe occupies the octahedral site (B-site) in the perovskite structure without having any magnetic cation as an immediate neighbour, this doublet is expected. The Raman and FTIR analysis of phase pure samples confirmed that $Ba^{2+}$ and $K^+$ go to the A-sites, whereas $Fe^{3+}$ and $Nb^{5+}$ occupy the B-sites of the $ABO_3$ perovskite structure of the BFN-KN solid solutions. Combining the XRD, FTIR, and Raman spectroscopy results, we observed that as the BFN content increases, the occupation of both Fe and Nb at the B-site (in place of only Nb) decreases the crystalline ordering than that is expected and observed for pure KN sample. Due to this gradual variation in structural ordering, the bandgap energies decrease as the BFN content increase in the solid solutions, resulting in a variation in the color of the corresponding samples. Through the analysis of the XRD, FTIR, Raman, UV-Vis, and SEM studies, we have correlated and introduced the concept that the structural disorder arising due to the position and size (strain) of the cations in the double perovskite structure are responsible for the typical frequency-dependent dielectric properties seen in BFN-KN solid solutions. Interestingly, our results confirm that there are three contributions to the interfacial polarization appearing at different frequencies. The disorder present at the grain boundary regions are quantitatively very high, and they give rise to the softer M-W interfacial polarization, whereas the atomic positional/chemical disorder, and the strain give rise to gradually harder M-W type interfacial polarizations, respectively. These results are corroborated with the optical and structural properties of our samples obtained through XRD, SEM, FTIR, UV-Vis, and Raman spectra. Our work adds to the fundamental understanding of dielectric polarization occurring in BFN-KN solid solutions having a double perovskite structure. The discussed mechanism can be used to explain many other similar systems too.

## Acknowledgment

BS acknowledges Dr. Emma Panzi (IISc.) for his help in the synthesis of the SSR samples and collecting the corresponding XRD data.

## Conflicts of interest

The authors declare no conflict of interest.

https://doi.org/10.1016/j.jiec.2018.07.006

57. Pascual-Gonzalez C, Schileo G, Feteira A. Band gap narrowing in ferroelectric KNbO$_3$-Bi(Yb,Me)O$_3$ (Me=Fe or Mn) ceramics. *Appl Phys Lett*. 2016;109(13):132902. https://doi.org/10.1063/1.4963699

58. Luisman L, Feteira A, Reichmann K. Weak-relaxor behaviour in Bi/Yb-doped KNbO$_3$ ceramics. *Appl Phys Lett*. 2011;99(192901):1–4. https://doi.org/10.1063/1.3660255

59. Quittet AM, Bell MI, Krauzman M, Raccah PM. Anomalous scattering and asymmetrical line shapes in Raman spectra of orthorhombic KNbO$_3$. *Phys Rev B*. 1976;14(11):5068–5072. https://doi.org/10.1103/PhysRevB.14.5068

60. Shen ZX, Hu ZP, Chong TC, Beh CY, Tang SH, Kuok MH. Pressure-induced strong mode coupling and phase transitions in KNbO$_3$. *Phys Rev B*. 1995;52(6):3976–3980. https://doi.org/10.1103/PhysRevB.52.3976

61. Bartasyte A, Kreisel J, Peng W, Guilloux-Viry M. Temperature-dependent Raman scattering of KTa$_{1-x}$Nb$_x$O$_3$ thin films. *Appl Phys Lett*. 2010;96(262903):1–3. https://doi.org/10.1063/1.3455326

62. Han F, Zhang Y, Yuan C, *et al.* Photocurrent and dielectric/ferroelectric properties of KNbO$_3$–BaFeO$_{3-\delta}$ ferroelectric semiconductors. *Ceram Int*. 2020;46(10):14567–14572. https://doi.org/10.1016/j.ceramint.2020.02.256

63. Siny IG, Katiyar RS, Bhalla AS. Cation Arrangement in the Complex Perovskites and Vibrational Spectra. *J Raman Spectrosc*. 1998;29(5):385–390. https://doi.org/10.1002/(sici)1097-4555(199805)29:5<385::aid-jrs250>3.3.co;2-6

64. Fujioka Y, Frantti J, Kakihana M. Raman scattering studies of the Ba$_2$MnWO$_6$ and Sr$_2$MnWO$_6$ double perovskites. *J Phys Chem B*. 2006;110(2):777–783. https://doi.org/10.1021/jp054861b

65. Zhang W, Wang Z, Chen XM. Crystal structure evolution and local symmetry of perovskite solid solution Ba(Fe$_{1/2}$Nb$_{1/2}$)$_{1-x}$Ti$_x$O$_3$ investigated by Raman spectra. *J Appl Phys*. 2011;110(6):64113. https://doi.org/10.1063/1.3639283

66. Correa M, Kumar A, Priya S, Katiyar RS, Scott JF. Phonon anomalies and phonon-spin coupling in oriented PbFe$_{0.5}$Nb$_{0.5}$O$_3$ thin films. *Phys Rev B*. 2011;83(1):14302.

Electronic Supporting Information

on

# Structure and dielectric properties of (x)BaFe$_{0.5}$Nb$_{0.5}$O$_3$ – (1-x)KNbO$_3$ solid solutions synthesized through solution combustion route

Vijay Khopkar and Balaram Sahoo[†]

*Materials Research Centre, Indian Institute of Science, Bangalore 560012 India*

[†]Corresponding author's email: bsahoo@iisc.ac.in (B. Sahoo)

Fig. S1(a) (below) shows the phase evolution of BaFe$_{0.5}$Nb$_{0.5}$O$_3$ (BFN) sample with sintering temperature. The XRD pattern of BFN sample immediate after solution combustion shows presence of many impurities and maybe some unreacted phases. For comparison, the ICSD data is also given. To get the phase pure BFN sample, the sample is calcined at elevated temperatures for 2 hours. It observed that for sample calcined at 1000 °C has small impurity peaks. As sintering temperature increased from 1100°C to 1200°C impurity peak disappeared. This shows the phase pure formation of BFN sample at calcination temperature of 1200°C. Similar observation was also made for rest of the solid solutions. The 0.8BFN-0.2KN form at 1100°C (Fig. S1(b)), 0.6BFN-0.4KN form at 1000 °C (Fig. S1(c)), 0.4BFN-0.6KN (Fig. S1(d)), and 0.2BFN-0.8KN (Fig S1(e)) form at 900 °C. It is interesting to see the XRD pattern of KNbO$_3$ (KN) sample, immediately after solution combustion method, whic seems to be forming even just after solution combustion reaction; however, it shows a very little amount of impurity phase, mostly carbonates. As calcination temperature increases this carbonate impurity peaks disappear, indicating the formation of phase pure KN sample. The KN sample is calcined at 800 °C to make it completely phase pure.

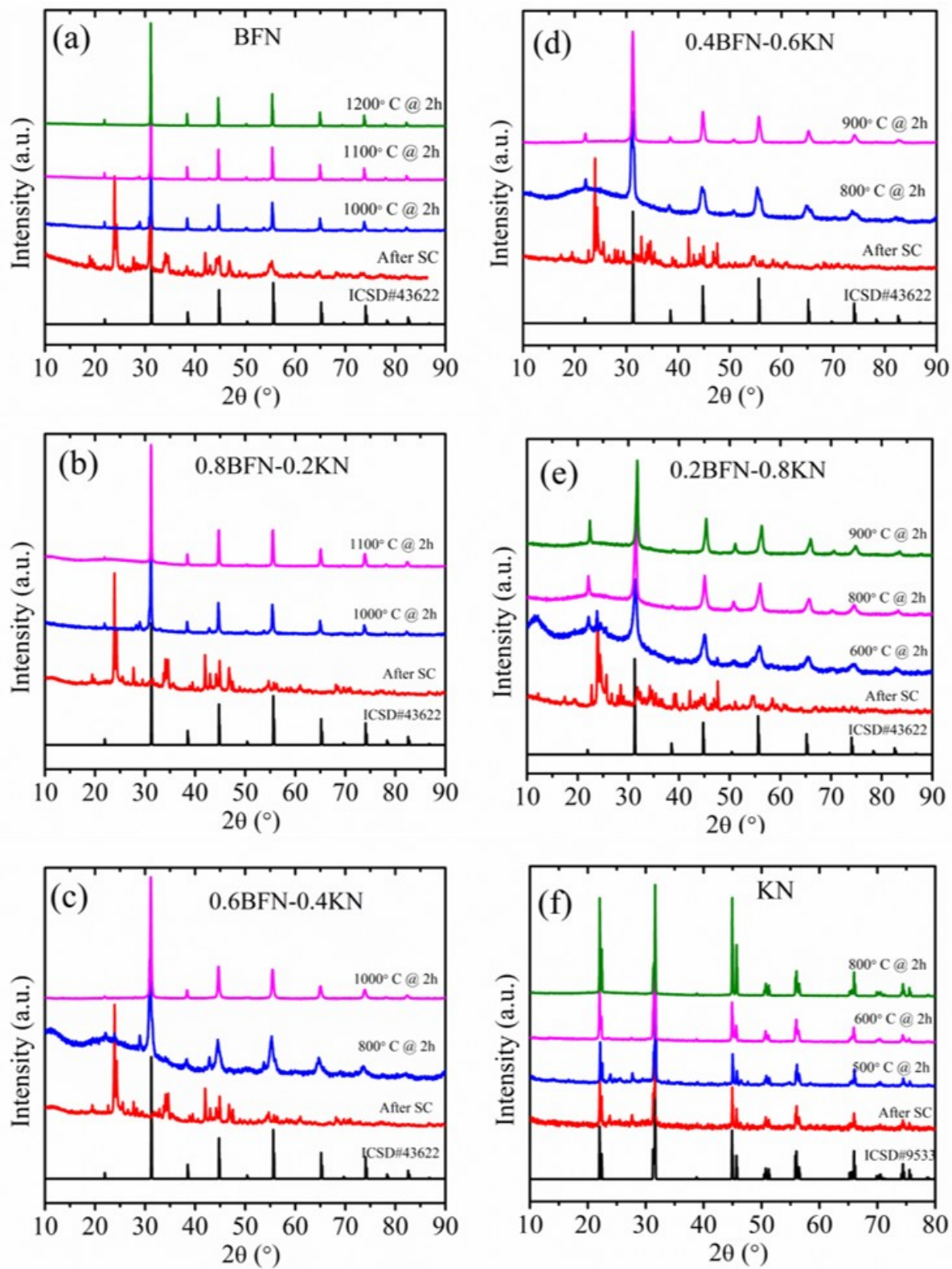

Fig. S1: (a) BFN, (b) 0.8BFN-0.2KN, (c) 0.6BFN-0.4KN, (d) 0.4BFN-0.6KN, (e) 0.2BFN-0.8KN, (f) KN, Phase evolution of solid solutions synthesized by solution combustion route at different elevated calcination temperature.

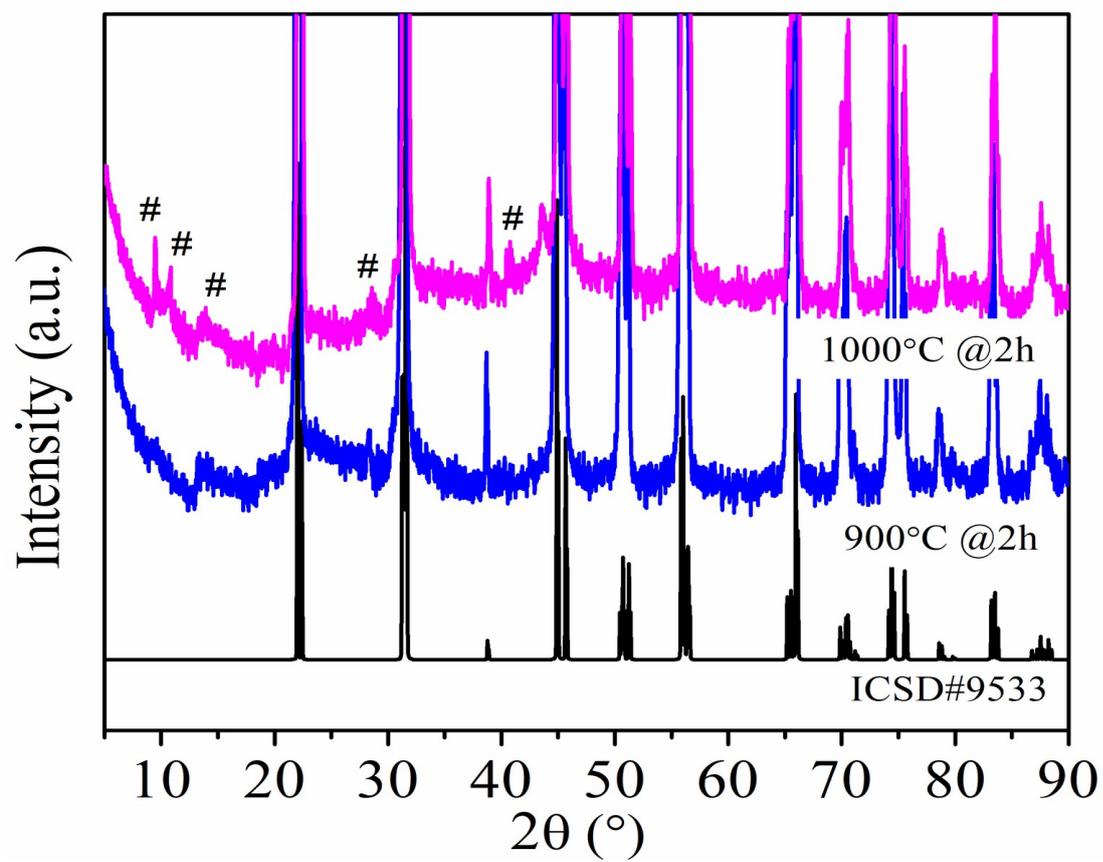

Fig. S2. The XRD patterns of the KN sample prepared by SCS method, after calcination at 900 and 1000 °C, as indicated. For comparison the plot of the ICSD file (# 9533) is also shown.

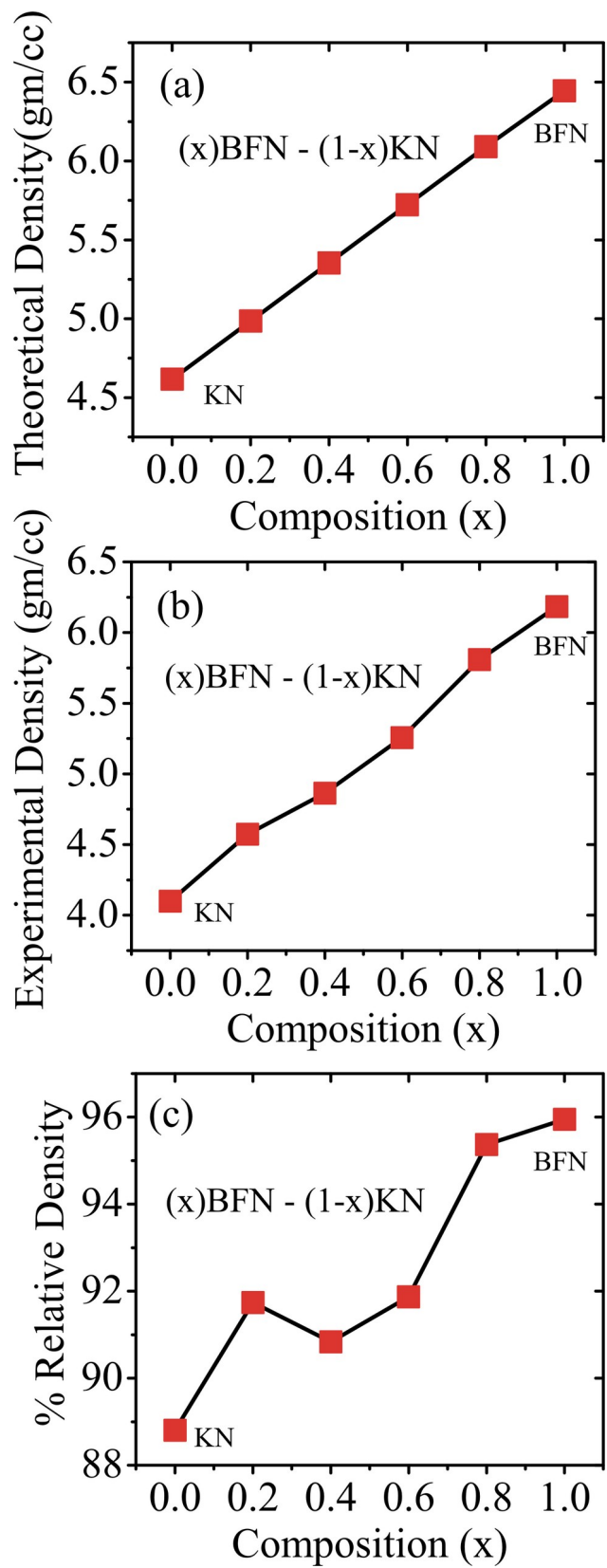

Fig. S3. (a): Theoretical density (b) experimental density (c) % Relative density of (x) BFN - (1-x) KN (x = 0, 0.2, 0.4, 0.6, 0.8, 1) solid solution pellets. The relative density suggests that there are lot of pores for the samples with low BFN content x.

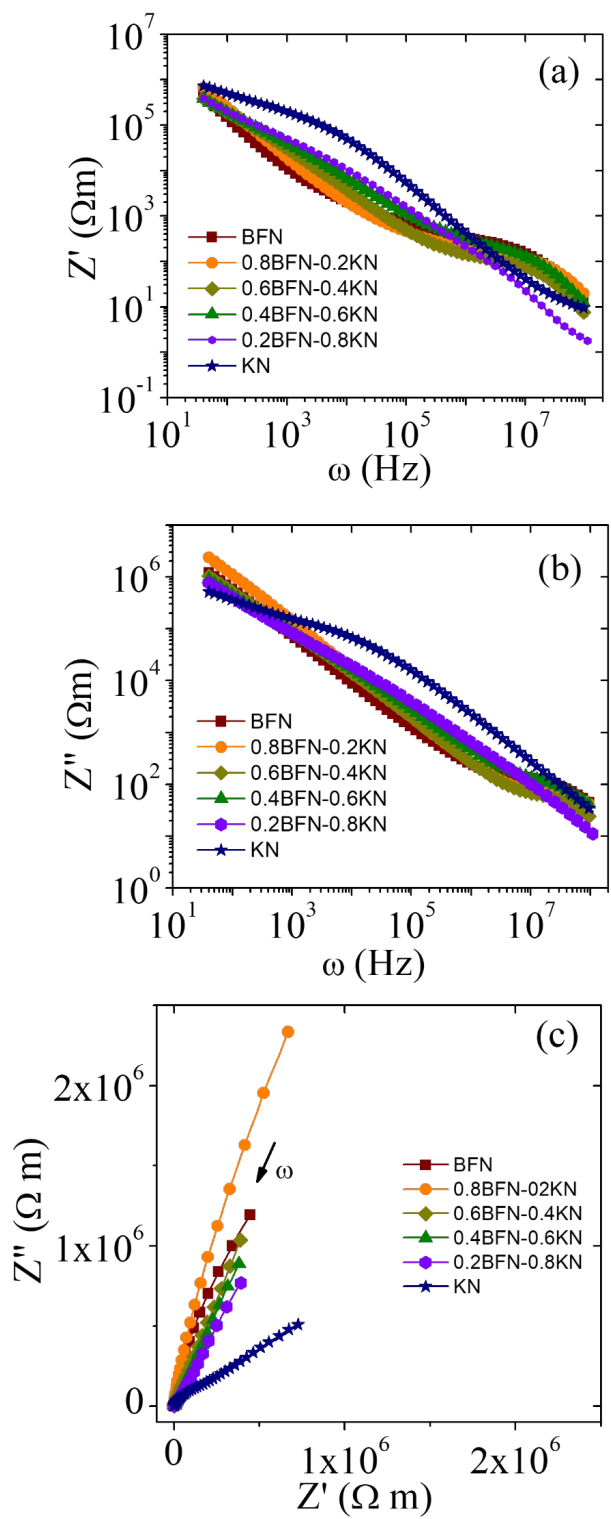

Fig. S4. (a) Real part of impedance, (b) imaginary part of impedance, (c) Cole-Cole plot for impedance of (x) BFN - (1-x) KN (x = 0, 0.2, 0.4, 0.6, 0.8, 1) solid solutions.